\documentclass[11pt]{article}
\usepackage[latin9]{inputenc}
\usepackage{amsmath}
\usepackage{amssymb}
\usepackage{graphicx}
\usepackage{esint}
\usepackage[unicode=true,
 bookmarks=false,
 breaklinks=false,pdfborder={0 0 1},backref=section,colorlinks=false]
 {hyperref}

\makeatletter

\DeclareFontEncoding{LGR}{}{}
\DeclareRobustCommand{\greektext}{%
  \fontencoding{LGR}\selectfont\def\encodingdefault{LGR}}
\DeclareRobustCommand{\textgreek}[1]{\leavevmode{\greektext #1}}
\ProvideTextCommand{\~}{LGR}[1]{\char126#1}

\providecommand{\tabularnewline}{\\}

\@ifundefined{date}{}{\date{}}

\usepackage{esint}
\usepackage{latexsym}
\usepackage{graphicx}\usepackage{bm}\usepackage{longtable}

\usepackage{xcolor}

\@addtoreset{equation}{section}

\makeatother

\begin{document}
\title{A holographic description of heavy-flavoured baryonic matter decay
involving glueball}
\maketitle
\begin{center}
\footnote{Email: siwenli@dlmu.edu.cn}Si-wen Li\emph{$^{\dagger}$}
\par\end{center}

\begin{center}
\emph{$^{\dagger}$Department of Physics, School of Science,}\\
\emph{Dalian Maritime University, }\\
\emph{Dalian 116026, China}\\
\par\end{center}

\vspace{8mm}

\begin{abstract}
We holographically investigate the decay of heavy-flavoured baryonic
hadron involving glueball by using the Witten-Sakai-Sugimoto model.
Since baryon in this model is recognized as the D4-brane wrapped on
$S^{4}$ and the glueball field is identified as the bulk gravitational
fluctuations, the interaction of the bulk graviton and the baryon
brane could be naturally interpreted as glueball-baryon interaction
through the holography which is nothing but the close-open string
interaction in string theory. In order to take account into the heavy
flavour, an extra pair of heavy-flavoured branes separated from the
other flavour branes with a heavy-light open string is embedded into
the bulk. Due to the finite separation of the flavour branes, the
heavy-light string creates massive multiplets which could be identified
as the heavy-light meson fields in this model. As the baryon brane
on the other hand could be equivalently described by the instanton
configuration on the flavour brane, we solve the equations of motion
for the heavy-light fields with the Belavin-Polyakov-Schwarz-Tyupkin
(BPST) instanton solution for the $N_{f}=2$ flavoured gauge fields.
Then with the solutions, we evaluate the soliton mass by deriving
the flavoured onshell action in strongly coupling limit and heavy
quark limit. After the collectivization and quantization, the quantum
mechanical system for glueball and heavy-flavoured baryon is obtained
in which the effective Hamiltonian is time-dependent. Finally we use
the standard technique for the time-dependent quantum mechanical system
to analyze the decay of heavy-flavoured baryon involving glueball
and we find one of the decay process might correspond to the decay
of baryonic B-meson involving the glueball candidate $f_{0}\left(1710\right)$.
This work is a holographic approach to study the decay of heavy-flavoured
hadron in nuclear physics.
\end{abstract}
\newpage{}

\section{Introduction}

Quantum Chromodynamics (QCD) as the fundamental theory of nuclear
physics predicts the bound state of pure gluons \cite{key-1,key-2,key-3}
because of its non-Abelian nature. Such bound state is always named
as ``glueball'' which is believed as the only possible composite
particle state in the pure Yang-Mills theory. In general glueball
states could have various Lorentz structures e.g. a scalar, pseudoscalar
or a tensor glueball with either normal or exotic $J^{PC}$ assignments.
Although the glueball state has not been confirmed by the experiment,
its spectrum has been studied by the simulation of lattice QCD \cite{key-4,key-5,key-6}.
According to the lattice calculations, it indicates that the lightest
glueball state is a scalar with assignment of $0^{++}$ and its mass
is around 1500-2000MeV \cite{key-4,key-7}. These results also suggest
that the scalar meson $f_{0}\left(1710\right)$ could be considered
as a glueball state. Glueball may be produced by the decay of various
hadrons in the heavy-ion collision \cite{key-8,key-9,key-10}, so
the dynamics of glueball is very significant. However lattice QCD
involving real-time quantities is extremely complexed and phenomenological
models usually include a large number of parameters with some corresponding
uncertainties. Thus it is still challenging to study the dynamics
of glueball with traditional quantum field theory.

Fortunately there is an alternatively different way to investigate
the dynamics of glueball based on the famous AdS/CFT correspondence
or the gauge-gravity duality pioneered in \cite{key-11} where a top-down
holographic approach from string theory by Witten \cite{key-12} and
Sakai and Sugimoto \cite{key-13} (i.e. the WSS model) is employed.
Analyzing the AdS/CFT dictionary with the WSS model, the glueball
field is identified as the bulk gravitational fluctuations carried
by the close strings while the meson states are created by the open
strings on the $N_{f}$ probe flavour branes. Hence this model naturally
includes the interaction of glueball and meson through the holography
which is nothing but the close-open string interaction in string theory.
Along this direction, decay of glueball into mesons has been widely
studied with this model e.g. in \cite{key-14,key-15,key-16}\footnote{Since the WSS model is based on $\mathrm{AdS_{7}/CFT_{6}}$ correspondence,
several previous work is also relevant to this model e.g. \cite{key-17,key-18}}.

\begin{figure}
\begin{centering}
\includegraphics[scale=0.35]{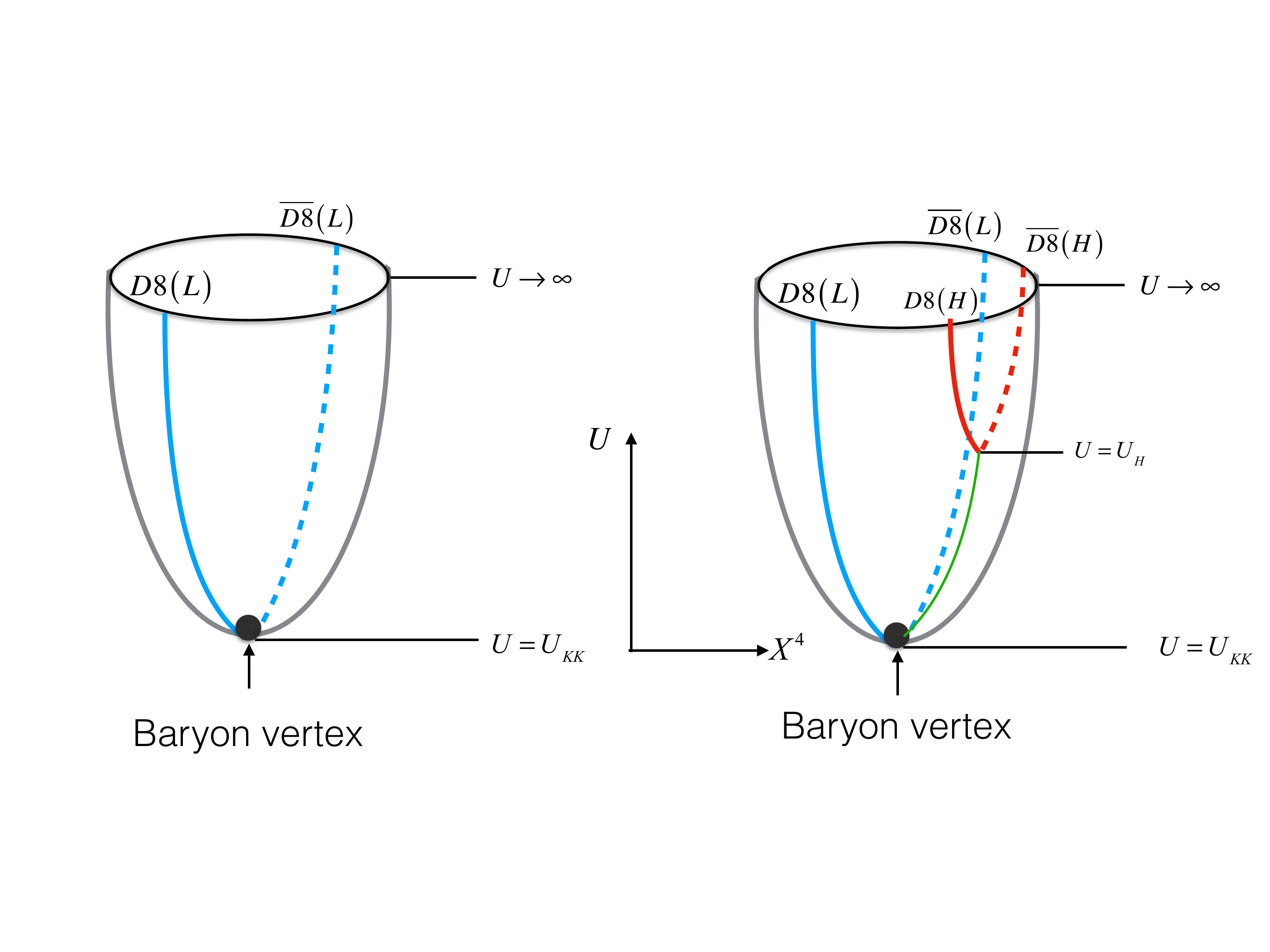}
\par\end{centering}
\caption{\label{fig:1}The various D-brane configurations in the WSS model.
\textbf{LEFT}: The configuration of the standard WSS model according
to Table \ref{tab:1}. The bulk geometry is produced by $N_{c}$ coincident
D4-branes which represent ``colours'' in QCD. The flavours are introduced
into the model by embedding $N_{f}$ pairs of coincident $\mathrm{D}8/\overline{\mathrm{D}8}$-branes
at the antipodal position of the bulk geometry. $U$ refers to the
holographic direction and $X^{4}$ is compactified on $S^{1}$. The
$\mathrm{D}4^{\prime}$-brane as the baryon vertex looks like a point
in the $U-X^{4}$ plane. Mesons are created by the open string on
the flavoured $\mathrm{D}8/\overline{\mathrm{D}8}$-branes while baryons
are created by the wrapped $\mathrm{D}4^{\prime}$-branes. \textbf{RIGHT}:
The WSS model with heavy flavour. An additional pair of flavoured
$\mathrm{D}8/\overline{\mathrm{D}8}$-brane (denoted by the red line)
as the heavy-flavoured (H) brane separated from the other $N_{f}$
pairs of light-flavoured (L) $\mathrm{D}8/\overline{\mathrm{D}8}$-branes
with a HL string (denoted by the green line) is embedded. The baryon
vertex contains heavy flavour in this configuration through the HL
string.}
\end{figure}

Keeping the above information in mind and partly motivated by \cite{key-8,key-9,key-10},
in this work we would like to holographically explore the glueball-baryon
interaction particularly involving the heavy flavour as an extension
to the previous study in \cite{key-19}. In the WSS model, baryon
is identified as the $\mathrm{D}4^{\prime}$-brane\footnote{We will use ``$\mathrm{D}4^{\prime}$-brane'' to distinguish the
baryon brane from those $N_{c}$ D4-branes as colour branes throughout
the manuscript.} wrapped on $S^{4}$ \cite{key-13,key-20} (namely the baryon vertex)
which could be equivalently described by the instanton configuration
on the flavoured D8-branes according to the string theory \cite{key-21,key-22}.
The configuration of various D-branes is illustrated in Table \ref{tab:1}.
In order to take account into the heavy flavour, we embed an extra
pair of flavoured $\mathrm{D}8/\overline{\mathrm{D}8}$-brane into
the bulk geometry which is separated from the other $N_{f}$ (light-flavoured)
$\mathrm{D}8/\overline{\mathrm{D}8}$-branes with an open string (the
heavy-light string) stretched between them \cite{key-23,key-24} as
illustrated in Figure \ref{fig:1}. In this configuration there would
be additional multiplets created by the heavy-light (HL) string and
they would acquire mass due to the finite separation of the flavour
branes. Hence we could interpret these multiplets as the HL meson
fields and the instanton configuration on the D8-branes with the multiplets
would include heavy flavour thus can be identified as heavy-flavoured
baryon \cite{key-25,key-26}. So similarly as the case of glueball
and meson, there must be glueball-baryon interaction in holography
as close string interacting with $\mathrm{D}4^{\prime}$-brane carrying
the heavy-flavour through the HL string, or namely graviton interacting
with the heavy-flavoured instantons.

\begin{table}[h]
\begin{centering}
\begin{tabular}{|c|c|c|c|c|c|c|c|c|c|c|}
\hline 
 & 0 & 1 & 2 & 3 & $\left(4\right)$ & 5$\left(U\right)$ & 6 & 7 & 8 & 9\tabularnewline
\hline 
\hline 
$\mathrm{Coloured}\ N_{c}\ \mathrm{D4}$  & - & - & - & - & - &  &  &  &  & \tabularnewline
\hline 
$\mathrm{Flavoured}\ N_{f}\ \mathrm{D}8/\overline{\mathrm{D}8}$ & - & - & - & - &  & - & - & - & - & -\tabularnewline
\hline 
Baryon vertex $\mathrm{D4}^{\prime}$ & - &  &  &  &  &  & - & - & - & -\tabularnewline
\hline 
\end{tabular}
\par\end{centering}
\caption{\label{tab:1}The brane configuration of the WSS model: \textquotedblleft -\textquotedblright{}
denotes the world volume directions of the D-branes.}
\end{table}

Let us outline the content and the organization of this manuscript
here. We consider the baryonic bound states created by the baryon
vertex in this model with two flavours  i.e. $N_{f}=2$. Following
\cite{key-22,key-24,key-25,key-26}, baryons are identified as Skyrmions
in the WSS model and they can be described by a quantum mechanical
system for their collective modes in the moduli space. The effective
Hamiltonian could be obtained by evaluating the classical mass of
the soliton $S^{\mathrm{onshel}}=-\int dtM_{soliton}$. So the main
goal of this paper is to evaluate the effective Hamiltonian involving
glueball-baryon interaction with heavy flavour. In section 2, we specify
the setup with the heavy flavour in this model and solve the classical
equations of motion for the HL meson field on the flavour brane. In
section 3, we search for the analytical solutions of the bulk gravitational
fluctuations then explicitly compute the onshell action with these
solutions. All the calculations are done in the limitation of large
't Hooft coupling constant $\lambda$ where the holography is exactly
valid. The final formulas of the effective Hamiltonian depend on the
glueball field and the number of heavy-flavoured quarks so that it
is time-dependent. Therefore the method for the time-dependent system
in quantum mechanics would be suitable to describe the decay of heavy-flavoured
baryons under the classical glueball field. Resultantly we obtain
several possible decay processes with the effective Hamiltonian and
pick out one of them which might probably correspond to the decay
of baryonic B-meson involving the glueball candidate $f_{0}\left(1710\right)$
as discussed in \cite{key-8,key-9,key-10}. 

Since the WSS model has been presented in many lectures, for reader's
convenience we only collect some relevant information about this model
in the Appendix A, B, C which can be also reviewed in \cite{key-13,key-22,key-25,key-26,key-27}.
Respectively the gravitational polarization used in this paper are
collected in Appendix A. Some useful formulas about the D-brane action
and the embedding of the probe branes and string in our setup can
be found in Appendix B. In Appendix C, it reviews the effective quantum
mechanical system for the collective modes of baryon. At the end of
this manuscript some messy but essential calculations about our main
discussion have been summarized in the Appendix D.

\section{Baryon as instanton with heavy flavour}

The baryon spectrum with pure light flavours in this model is reviewed
in the Appendix C, so we only outline how to include the heavy flavour
in this section. Some necessary information about the embedding of
the probe branes and string in our setup could be reviewed in Appendix
B.

A simple way to involve the heavy flavour in this model is to embed
an extra pair of flavoured $\mathrm{D}8/\overline{\mathrm{D}8}$-brane
separated from the other $N_{f}$ (light-flavoured) $\mathrm{D}8/\overline{\mathrm{D}8}$-branes
with an open string (the heavy-light string) stretched between them
as illustrated in Figure \ref{fig:1}. The HL string creates additional
multiplets according to the string theory \cite{key-27} since it
connects to the separated branes. And these multiplets could be approximated
by local vector fields near the worldvolume of the light flavour branes.
Note that the multiplets acquire mass due to the finite length, or
namely the non-zero vacuum expectation value (VEV) of the HL string.
Therefore we could interpret the multiplets created by the HL string
as the heavy-flavoured mesons with massive flavoured (heavy-flavoured)
quarks. Actually this mechanism to acquire mass is nothing but the
``Higgs mechanism'' in string theory. So let us replace the gauge
fields on the light flavour branes by its matrix-valued form to involve
the heavy flavour,

\begin{equation}
\mathcal{A}_{a}\rightarrow\boldsymbol{\mathrm{\mathcal{A}}}_{a}=\left(\begin{array}{cc}
\mathcal{A}_{a} & \Phi_{a}\\
-\Phi_{a}^{\dagger} & 0
\end{array}\right).\label{eq:1}
\end{equation}
In our notation $\mathcal{A}_{a}$ is an $N_{f}\times N_{f}$ matrix-valued
1-form while $\boldsymbol{\mathrm{\mathcal{A}}}_{a}$ is an $\left(N_{f}+1\right)\times\left(N_{f}+1\right)$
matrix-valued 1-form. $\Phi_{a}$ is an $N_{f}\times1$ matrix-valued
vector which represents HL meson field and the index runs over the
light flavour brane. Thus the field strength of $\boldsymbol{\mathrm{\mathcal{A}}}_{a}$
also becomes matrix-valued as a 2-form,

\begin{equation}
\mathcal{F}_{ab}\rightarrow\boldsymbol{\mathrm{\mathcal{F}}}_{ab}=\left(\begin{array}{cc}
\mathcal{F}_{ab}-\Phi_{[a}\Phi_{b]}^{\dagger} & \partial_{[a}\Phi_{b]}+\mathcal{A}_{[a}\Phi_{b]}\\
-\partial_{[a}\Phi_{b]}^{\dagger}-\Phi_{[a}^{\dagger}\mathcal{A}_{b]} & -\Phi_{[a}^{\dagger}\Phi_{b]}
\end{array}\right),\label{eq:2}
\end{equation}
where $\mathcal{F}_{ab}$ refers to the field strength of $\mathcal{A}_{a}$.
Imposing (\ref{eq:1}) (\ref{eq:2}) into D8-brane action (\ref{eq:C-1})
and keep the quadric terms of $\Phi_{a}$, it leads to a Yang-Mills
(YM) action\footnote{We do not given the explicit formula of the CS term with HL fields
since it is independent on the metric thus it is irrelevant to the
glueball-baryon interaction.}

\begin{align}
S_{\mathrm{DBI}}^{\mathrm{YM}}= & -\frac{1}{4}\left(2\pi\alpha^{\prime}\right)^{2}T_{8}\mathrm{Tr}\int_{\mathrm{D}8/\overline{\mathrm{D}8}}d^{9}xe^{-\Phi}\sqrt{-g}g^{ab}g^{cd}\boldsymbol{\mathcal{F}}{}_{ac}\boldsymbol{\mathcal{F}}_{bd}\nonumber \\
= & -\frac{1}{4}\left(2\pi\alpha^{\prime}\right)^{2}T_{8}\int_{\mathrm{D}8/\overline{\mathrm{D}8}}d^{9}xe^{-\Phi}\sqrt{-g}\bigg[g^{ab}g^{cd}\mathrm{Tr}\left(\mathcal{F}_{ac}\mathcal{F}_{bd}-\alpha_{ac}\mathcal{F}_{bd}-\mathcal{F}_{ac}\alpha_{bd}\right)\nonumber \\
 & -2g^{ab}g^{cd}f_{ac}^{\dagger}f_{bd}\bigg],\label{eq:3}
\end{align}
where 

\begin{equation}
f_{ab}=\partial_{[a}\Phi_{b]}+A_{[a}\Phi_{b]},\ f_{ab}^{\dagger}=-\partial_{[a}\Phi_{b]}^{\dagger}-\Phi_{[a}^{\dagger}A_{b]},\ \alpha_{ab}=\Phi_{[a}\Phi_{b]}^{\dagger}.
\end{equation}
We should notice that from the full formula of the DBI action, it
would contain an additional term of the transverse mode $\Psi$ of
$\mathrm{D}8/\overline{\mathrm{D}8}$-branes as shown in Appendix
B. And this term could be written as,

\begin{equation}
S_{\Psi}^{\mathrm{D8}}=-\tilde{T}_{8}\mathrm{Tr}\int_{\mathrm{D}8/\overline{\mathrm{D}8}}d^{9}xe^{-\Phi}\sqrt{-\det g}\left\{ \frac{1}{2}D_{a}\Psi D^{a}\Psi+\frac{1}{4}\left[\Psi,\Psi\right]^{2}\right\} ,\label{eq:5}
\end{equation}
with $D_{a}\Psi=\partial_{a}\Psi+\left[\boldsymbol{\mathrm{A}}_{a},\Psi\right]$
and $\tilde{T}_{8}=\left(2\pi\alpha^{\prime}\right)^{2}T_{8}$. In
the case of $N_{f}$ pairs of light-flavoured $\mathrm{D}8/\overline{\mathrm{D}8}$
branes separated from one pair of heavy-flavoured $\mathrm{D}8/\overline{\mathrm{D}8}$
brane, we can define the moduli solution of $\Psi$ with a finite
VEV $v$ by the extrema of the potential contribution or $\left[\Psi,\left[\Psi,\Psi\right]\right]=0$
\cite{key-27,key-28} as,

\begin{equation}
\Psi=\left(\begin{array}{cc}
-\frac{v}{N_{f}}\boldsymbol{1}_{N_{f}} & 0\\
0 & v
\end{array}\right).\label{eq:6}
\end{equation}
So the action (\ref{eq:5}) could be rewritten by plugging the solution
(\ref{eq:6}) into (\ref{eq:5}) as,

\begin{equation}
S_{\Psi}=-\tilde{T}_{8}v^{2}\frac{\left(N_{f}+1\right)^{2}}{N_{f}^{2}}\mathrm{Tr}\int d^{4}x\int_{-\infty}^{+\infty}dZe^{-\Phi}\sqrt{-\det g}g^{ab}\Phi_{a}^{\dagger}\Phi_{b},\label{eq:7}
\end{equation}
which is exactly the mass term of the HL field. Then perform the rescaling
(\ref{eq:C-3}), we could obtain the classical equations of motions
for $\Phi_{a}$ from (\ref{eq:3}) (\ref{eq:7}) as,

\begin{align}
D_{M}D_{M}\Phi_{N}-D_{N}D_{M}\Phi_{M}+2\mathcal{F}_{NM}\Phi_{M}+\mathcal{O}\left(\lambda^{-1}\right) & =0.\nonumber \\
D_{M}\left(D_{0}\Phi_{M}-D_{M}\Phi_{0}\right)-\mathcal{F}^{0M}\Phi_{M}-\frac{1}{64\pi^{2}a}\epsilon_{MNPQ}K_{MNPQ}+\mathcal{O}\left(\lambda^{-1}\right) & =0,\label{eq:8}
\end{align}
where $x^{M}=\left\{ x^{i},Z\right\} ,i=1,2,3$ and the 4-form $K_{MNPQ}$
is given as,

\begin{equation}
K_{MNPQ}=\partial_{M}\mathcal{A}_{N}\partial_{P}\Phi_{Q}+\mathcal{A}_{M}\mathcal{A}_{N}\partial_{P}\Phi_{Q}+\partial_{M}\mathcal{A}_{N}\mathcal{A}_{P}\Phi_{Q}+\frac{5}{6}\mathcal{A}_{M}\mathcal{A}_{N}\mathcal{A}_{P}\Phi_{Q}.
\end{equation}
Since the holographic approach is valid in the strongly coupling limit
$\lambda\rightarrow\infty$, the contributions from $\mathcal{O}\left(\lambda^{-1}\right)$
have been dropped off. Note that the light flavoured gauge field $\mathcal{A}_{a}$
satisfies the equations of motion obtained by varying the action (\ref{eq:C-1}),
so their solution remains to be (\ref{eq:C-2}) in the large $\lambda$
limit. And we could further define $\Phi_{a}=\phi_{a}e^{\pm im_{H}x^{0}}$
in the heavy quark limit i.e. $m_{H}\rightarrow\infty$ as in \cite{key-25,key-26}
so that $D_{0}\Phi_{M}=\left(D_{0}\pm im_{H}\right)\phi_{M}$ where
``$\pm$'' corresponds to quark and anti-quark respectively. By
keeping these in mind, altogether we find the full solution for (\ref{eq:8})
as,

\begin{align}
\phi_{0} & =-\frac{1}{1024a\pi^{2}}\left[\frac{25\rho}{2\left(x^{2}+\rho^{2}\right)^{5/2}}+\frac{7}{\rho\left(x^{2}+\rho^{2}\right)^{3/2}}\right]\chi,\nonumber \\
\phi_{M} & =\frac{\rho}{\left(x^{2}+\rho^{2}\right)^{3/2}}\sigma_{M}\chi,\label{eq:10}
\end{align}
where $\chi$ is a spinor independent on $x^{M}$. Then in the double
limit i.e. $\lambda\rightarrow\infty$ followed by $m_{H}\rightarrow\infty$,
the Hamiltonian for the collective modes involving the heavy flavour
could be calculated as in (\ref{eq:C-7}) by following the procedures
in Appendix C. 

\section{Glueball-baryon interaction with heavy flavour}

The dynamic of free glueball is reviewed in Appendix A, so in this
section we will focus on the interaction of glueball and baryon with
heavy flavour charactered by the collective Hamiltonian. As the glueball
field is included by the metric fluctuations, the Chern-Simons (CS)
term is independent on the metric thus it does not involve the glueball-baryon
interaction. Hence let us start with the five dimensional (5d) YM
action plus the mass term which are collected in (\ref{eq:3}) (\ref{eq:7}).
The onshell form of (\ref{eq:3}) (\ref{eq:7}) corresponds to the
effective interaction Hamiltonian of glueball and heavy-flavoured
baryon through the relation $S^{\mathrm{onshel}}=-\int dtH_{G-B}$,
accordingly we first need to solve the eigenvalue equations for function
$H_{E,D,T}$ in large $\lambda$ limit.

The eigenvalue equations for $H_{E,D,T}$ are given in (\ref{eq:A-9})
and (\ref{eq:A-14}). In the rescaled coordinate $Z\rightarrow\lambda^{-1/2}Z$,
the equations are written as,

\begin{align}
H_{E}^{\prime\prime}\left(Z\right)+\left(\frac{1}{Z}+\frac{2Z}{\lambda}\right)H_{E}^{\prime}\left(Z\right)+\left(\frac{16}{3\lambda}+\frac{M_{E}^{2}}{M_{KK}^{2}}\frac{1}{\lambda}\right)H_{D}\left(Z\right)+\mathcal{O}\left(\lambda^{-2}\right) & =0,\nonumber \\
H_{D,T}^{\prime\prime}\left(Z\right)+\left(\frac{1}{Z}+\frac{2Z}{\lambda}\right)H_{D,T}^{\prime}\left(Z\right)+\frac{M_{D}^{2}}{M_{KK}^{2}}\frac{1}{\lambda}H_{D,T}\left(Z\right)+\mathcal{O}\left(\lambda^{-2}\right) & =0,
\end{align}
and they could be easily solved as,

\begin{align}
H_{E}\left(z\right) & =\mathcal{C}_{E}\left(1-\frac{3M_{E}^{2}+16M_{KK}^{2}}{12M_{KK}^{2}\lambda}Z^{2}\right)\lambda^{-1/2}N_{c}^{-1}M_{KK}^{-1}+\mathcal{O}\left(\lambda^{-3/2}\right),\nonumber \\
H_{D,T}\left(z\right) & =\mathcal{C}_{D,T}\left(1-\frac{M_{D,T}^{2}}{4M_{KK}^{2}\lambda}Z^{2}\right)\lambda^{-1/2}N_{c}^{-1}M_{KK}^{-1}+\mathcal{O}\left(\lambda^{-3/2}\right).\label{eq:12}
\end{align}

Next we perform the rescaling as in (\ref{eq:C-3}), then insert the
BPST solution (\ref{eq:C-2}) for the gauge field $\mathcal{A}$ and
(\ref{eq:A-9}) for the heavy-light meson field $\Phi_{a}$ into the
action (\ref{eq:3}) (\ref{eq:7}). Afterwards by plugging the metric
(\ref{eq:A-6}) plus the dilaton (\ref{eq:A-7}) with the solution
(\ref{eq:12}) and various fluctuations which include the exotic scalar,
dilatonic scalar and tensor glueball field all given in the Appendix
A into the action (\ref{eq:3}) (\ref{eq:7}), the onshell form of
action (\ref{eq:3}) (\ref{eq:7}) could be obtained with the dimensionless
variable $x^{\mu}\rightarrow x^{\mu}/M_{KK},\boldsymbol{\mathcal{A}}_{\mu}\rightarrow\boldsymbol{\mathcal{A}}_{\mu}M_{KK}$
as,

\begin{align}
S_{G_{E,D,T}-B}^{\mathrm{onshell}}= & a\mathcal{C}_{E,D,T}\int d^{4}xdZ\mathrm{Tr}\bigg[\lambda^{1/2}\mathcal{L}_{1/2}^{E,D,T}+\mathcal{L}_{0}^{E,D,T}+\lambda^{-1/2}\mathcal{L}_{-1/2}^{E,D,T}+\lambda^{-1/2}\mathcal{L}_{\Psi}^{E,D,T}\nonumber \\
 & +\mathcal{O}\left(\lambda^{-1}m_{H}^{0}\right)\bigg],\label{eq:13}
\end{align}
where $a=\frac{1}{216\pi^{3}}$, ``E,D,T'' refers respectively to
``exotic scalar, dilatonic scalar and tensor glueball''. Although
the above calculation is very straightforward, the explicit forms
of $\mathcal{L}_{1/2,0,-1/2}^{E,D,T}$ and $\mathcal{L}_{\Psi}^{E,D,T}$
are quite lengthy. So we summarize the full formulas of $\mathcal{L}_{1/2,0,-1/2}^{E,D,T}$
and $\mathcal{L}_{\Psi}^{E,D,T}$ with some essential instructions
in Appendix D and here skip to the final results. Using the relation
$S_{G-B}^{\mathrm{onshell}}=-\int dtH_{G-B}\left(t,\mathcal{X}^{s}\right)$,
 the dimensionless interaction Hamiltonians are computed as\footnote{The glueball field $G_{E,D,T}$ in (\ref{eq:14}) is dimensional which
is in the unit of $M_{KK}$ while the other parameters are dimensionless.},

\begin{align}
H_{G_{E}-B}\left(t,\mathcal{X}^{s}\right)= & -\mathcal{C}_{E}\lambda^{-1/2}M_{KK}^{-1}\left(5m_{H}^{2}\pi^{2}a+\frac{15m_{H}}{32\rho^{2}}\right)G_{E}\chi^{\dagger}\chi+\mathcal{O}\left(\lambda^{-1}m_{H}^{0}\right)\nonumber \\
H_{G_{D}-B}\left(t,\mathcal{X}^{s}\right)= & \mathcal{C}_{D}\lambda^{-1/2}M_{KK}^{-1}\left(\frac{3m_{H}}{8\rho^{2}}-6m_{H}^{2}\pi^{2}a\right)G_{D}\chi^{\dagger}\chi+\mathcal{O}\left(\lambda^{-1}m_{H}^{0}\right)\nonumber \\
H_{G_{T}-B}\left(t,\mathcal{X}^{s}\right)= & -\mathcal{C}_{T}\lambda^{-1/2}M_{KK}^{-1}\left(\frac{7}{2}m_{H}^{2}\pi^{2}a+\frac{m_{H}}{4\rho^{2}}\right)G_{T}\chi^{\dagger}\chi+\mathcal{O}\left(\lambda^{-1}m_{H}^{0}\right),\label{eq:14}
\end{align}
The constants $\mathcal{C}_{E,D,T}$ are determined by the eigenvalue
equations for $H_{E,D,T}$ and they depends on the mass of the various
glueballs. We numerically evaluate $\mathcal{C}_{E,D,T}$ in Table
\ref{tab:2} with the corresponding glueball mass. Notice that the
operator $G_{E,D,T}$ satisfies the equations of motion by varying
action (\ref{eq:A-10}) (\ref{eq:A-15}), thus its classical solution
is $G_{E,D,T}=\frac{1}{2}\left(e^{-iM_{E,D,T}t}+\mathrm{c.c}\right)$
and it is time-dependent. On the other hand, the spinor $\chi$ has
to be however quantized by its anti-commutation relation $\left\{ \chi_{\alpha},\chi_{\beta}^{\dagger}\right\} =\delta_{\alpha\beta}$
in the full quantum field theory so $\chi^{\dagger}\chi$ is the number
operator of heavy quarks. Therefore in our theory the glueball field
could be treated as the classical field while baryon is quantized
in the moduli space and we can identify $\chi^{\dagger}\chi=N_{Q}$
as the number of heavy quarks in a baryonic bound state. Moreover
the Hamiltonians in (\ref{eq:14}) is definitely suitable to be perturbations
to the quantum mechanics (\ref{eq:C-7}) since they are all proportional
to $\lambda^{-1/2}$ in the large $\lambda$ limit. Then the interaction
of glueball and heavy-flavoured baryon could be accordingly described
by using the method of time-dependent perturbation in the quantum
mechanical system. Last but not least, the decay rates/width $\Gamma$
can be evaluated by using the standard technique for the time-dependent
perturbation in quantum mechanics, which is given as,

\begin{align}
\frac{\Gamma_{B\rightarrow G+X}}{m_{H}}= & \frac{1}{m_{H}}\left|\int dt\left\langle i\left|H_{G-B}\left(t,\mathcal{X}^{s}\right)\right|j\right\rangle e^{-i\left(E_{i}-E_{j}\right)t}\right|^{2},\nonumber \\
= & \frac{1}{m_{H}}\left\langle i\right|H_{G-B}\left(\mathcal{X}^{s}\right)\left|j\right\rangle ^{2}\delta\left(E_{j}-E_{i}-M_{E,D,T}\right),\label{eq:15}
\end{align}
$\left|i\right\rangle ,\left|j\right\rangle ,E_{i,j}$ refers to the
eigenstate and the associated eigenvalue of (\ref{eq:C-7}). And the
above decay occurs only if several physical quantities e.g. energy,
total angular momentum $J$, are also conserved. Note that the interaction
Hamiltonians in (\ref{eq:14}) are independent on $Z$, so $\left\langle i\right|H_{G-B}\left(\mathcal{X}^{s}\right)\left|j\right\rangle $
would be vanished unless the states $\left|i\right\rangle ,\left|j\right\rangle $
take the same quantum number of $n_{Z}$ and $l$. The Hamiltonians
in (\ref{eq:14}) can also describe the decay of an anti-baryon if
we replace $m_{H}$ by $-m_{H}$.

\begin{table}
\begin{centering}
\begin{tabular}{|c|c|c|c|c|c|}
\hline 
Excitation of glueball $\left(n\right)$ & $n=0$ & $n=1$ & $n=2$ & $n=3$ & $n=4$\tabularnewline
\hline 
Glueball mass $M_{E}^{\left(n\right)}$ & 0.901 & 2.285 & 3.240 & 4.149 & 5.041\tabularnewline
\hline 
Glueball mass $M_{D,T}^{\left(n\right)}$ & 1.567 & 2.485 & 3.373 & 4.252 & 5.124\tabularnewline
\hline 
\hline 
The coefficients & $n=0$ & $n=1$ & $n=2$ & $n=3$ & $n=4$\tabularnewline
\hline 
$\mathcal{C}_{E}$ & 144.545 & 114.871 & 131.283 & 146.259 & 157.832\tabularnewline
\hline 
$\mathcal{C}_{D}$ & 29.772 & 36.583 & 42.237 & 47.220 & 51.724\tabularnewline
\hline 
$\mathcal{C}_{T}$ & 72.927 & 89.609 & 103.46 & 115.664 & 126.696\tabularnewline
\hline 
\end{tabular}
\par\end{centering}
\caption{\label{tab:2}The glueball mass spectrum $M_{E,T}^{\left(n\right)}$
in the WSS model in the units of $M_{KK}$ is collected from \cite{key-14}
and the numerical values of the associated coefficients presented
in (\ref{eq:14}) $\mathcal{C}_{E,D,T}$ are evaluated.}
\end{table}

With the perturbed Hamiltonian in (\ref{eq:14}), this model includes
various decays of heavy-flavoured hadrons involving the glueball.
So we are going to examine the possible transitions involving one
glueball with the leading low-energy excited baryon states $n_{\rho}\leq5$.
Since our concern is the situation of two-flavoured meson, we could
follow \cite{key-22} by setting $\frac{l}{2}=J=0,N_{Q}=1,N_{c}=3$
in order to fit the experimental data of the (pseudo) scalar meson
states with one heavy flavour. Then let us first take account into
the energy conservation $E\left(n_{\rho}=n_{\rho}^{\prime}+\Delta n_{\rho},l=0,N_{B}=1,n_{Z}\right)-E\left(n_{\rho}^{\prime},l=0,N_{B}=1,n_{Z}\right)\equiv\mathcal{E}\left(\Delta n_{\rho}\right)=M_{E,D,T}^{\left(n\right)}$
if the transition of hadron decay would happen, where $M_{E,D,T}^{\left(n\right)}$
refers to the glueball mass given in Table \ref{tab:2} and $E\left(n_{\rho},l,N_{B},n_{Z}\right)$
refers to the baryonic spectrum in (\ref{eq:C-9}). By keeping these
in mind, the following relations are picked out,

\begin{equation}
\mathcal{E}\left(\Delta n_{\rho}=3\right)/M_{D,T}^{\left(n=1\right)}\simeq0.986,\ \mathcal{E}\left(\Delta n_{\rho}=4\right)/M_{E}^{\left(n=2\right)}\simeq1.008,\label{eq:16}
\end{equation}
while $\mathcal{E}\left(\Delta n_{\rho}\right),n_{\rho}\leq5$ with
$\Delta n_{\rho}=0,1,2$ does not match to any $M_{E,D,T}^{\left(n\right)}$.
Hence we could find the following possible decays involving glueball
according to (\ref{eq:16}),

\begin{align}
\mathrm{I:} & Baryonic\left|J=0,n_{\rho}=3\right\rangle \rightarrow\left|G_{D}^{\left(n=1\right)},J^{PC}=0^{++}\right\rangle +Baryonic\left|J=0,n_{\rho}=0\right\rangle \nonumber \\
\mathrm{II:} & Baryonic\left|J=0,n_{\rho}=4\right\rangle \rightarrow\left|G_{D}^{\left(n=1\right)},J^{PC}=0^{++}\right\rangle +Baryonic\left|J=0,n_{\rho}=1\right\rangle \nonumber \\
\mathrm{III:} & Baryonic\left|J=0,n_{\rho}=5\right\rangle \rightarrow\left|G_{D}^{\left(n=1\right)},J^{PC}=0^{++}\right\rangle +Baryonic\left|J=0,n_{\rho}=2\right\rangle \nonumber \\
\mathrm{IV:} & Baryonic\left|J=0,n_{\rho}=3\right\rangle \rightarrow\left|G_{T}^{\left(n=1\right)},J^{PC}=2^{++}\right\rangle +Baryonic\left|J=0,n_{\rho}=0\right\rangle \nonumber \\
\mathrm{V:} & Baryonic\left|J=0,n_{\rho}=4\right\rangle \rightarrow\left|G_{T}^{\left(n=1\right)},J^{PC}=2^{++}\right\rangle +Baryonic\left|J=0,n_{\rho}=1\right\rangle \nonumber \\
\mathrm{VI:} & Baryonic\left|J=0,n_{\rho}=5\right\rangle \rightarrow\left|G_{T}^{\left(n=1\right)},J^{PC}=2^{++}\right\rangle +Baryonic\left|J=0,n_{\rho}=2\right\rangle \nonumber \\
\mathrm{VII:} & Baryonic\left|J=0,n_{\rho}=4\right\rangle \rightarrow\left|G_{E}^{\left(n=2\right)},J^{PC}=0^{++}\right\rangle +Baryonic\left|J=0,n_{\rho}=0\right\rangle \nonumber \\
\mathrm{VIII}: & Baryonic\left|J=0,n_{\rho}=5\right\rangle \rightarrow\left|G_{E}^{\left(n=2\right)},J^{PC}=0^{++}\right\rangle +Baryonic\left|J=0,n_{\rho}=1\right\rangle ,\label{eq:17}
\end{align}
where we have denoted the states by their quantum numbers and the
associated decay rates $\Gamma$ are numerically evaluated in Table
\ref{tab:3} by using the effective Hamiltonian in (\ref{eq:14}).
Notice that the mass of the dilatonic and exotic scalar glueball in
(\ref{eq:17}) are given as $M_{E}^{\left(n=2\right)}/M_{D}^{\left(n=1\right)}\simeq1.30$
which is close to the mass ratio of the glueball candidates $f_{0}\left(1710\right)$
and $f_{0}\left(1500\right)$ as $M_{f_{0}\left(1710\right)}/M_{f_{0}\left(1500\right)}\simeq1.14$,
moreover all of them should be the state of $J^{PC}=0^{++}$. Accordingly
we could identify the dilatonic and exotic scalar glueball in (\ref{eq:17})
to $f_{0}\left(1500\right)$ and $f_{0}\left(1710\right)$ respectively
which are the two glueball candidates discussed frequently in many
lectures.

\begin{table}[h]
\begin{centering}
\begin{tabular}{|c|c|c|c|c|}
\hline 
 & I & II & III & IV\tabularnewline
\hline 
$\Gamma$ & 0.0392$\lambda^{-1}$ & 0.0628$\lambda^{-1}$ & 0.0785$\lambda^{-1}$ & 0.1046$\lambda^{-1}$\tabularnewline
\hline 
\hline 
 & V & VI & VII & VIII\tabularnewline
\hline 
$\Gamma$ & 0.1674$\lambda^{-1}$ & 0.2093$\lambda^{-1}$ & 0.6316$\lambda^{-1}$ & 1.0527$\lambda^{-1}$\tabularnewline
\hline 
\end{tabular}
\par\end{centering}
\caption{\label{tab:3} The corresponding decay rates in the units of $m_{H}$
to the transitions in (\ref{eq:17}) by setting $l=0,N_{Q}=1,N_{c}=3,N_{f}=2$.}
\end{table}

If we furthermore consider the parity of baryonic states as discussed
in \cite{key-22}, the above states with odd $n_{Z}$ in this model
would correspond to the meson states with odd parity since the parity
transformation is $Z\rightarrow-Z$. In this sense, the transition
II, V, VII describes the decay of the heavy-flavoured scalar (non-glueball)
meson involving glueball while the pure scalar meson with even parity
is less evident according to the current experimental data. On the
other hand, as the glueball states we discussed in this manuscript
all have even parity, it implies that the parity of the transition
I, III, IV, VI may be violated. We also notice that if $\frac{l}{2}=J$
is identified as the quantum number of the spin, the decay processes
IV, V, VI in (\ref{eq:17}) involving tensor glueball $J^{PC}=2^{++}$
may be probably forbidden since the initial and final baryonic states
are all pure scalars i.e. the total angular momentum may not be conserved
in these transitions \footnote{For a tensor glueball, we suggest to consider a tensor field dependent
on the coordinates of the moduli space $y_{I}$ in order to obtain
the correct decay process. We would like to leave it as a future study
and focus on the scalar glueball in the current work.}, and this result would be in agreement with the previous discussion
in \cite{key-19}. Therefore we could conclude that only the decay
process VIII in (\ref{eq:17}) might be realistic. This transition
describes the decay of the baryonic meson consisted of one heavy-
and one light- flavoured quark. So while the identification of the
other transitions might be less clear, the transition VIII could be
interpreted as the decay of the baryonic B-meson involving the glueball
candidate $f_{0}\left(1710\right)$ as discussed e.g. in \cite{key-8,key-9,key-10}
since the corresponding quantum numbers of the states could be identified.

\section{Summary}

In this paper, with the top-down approach of WSS model, we propose
a holographic description of the decay of heavy-flavoured meson involving
glueball. The HL field is introduced into the WSS model to describe
the dynamics of heavy flavour and it is created by the HL string with
a pair of heavy-flavoured $\mathrm{D}8/\overline{\mathrm{D}8}$-brane
separated from the other light flavoured $\mathrm{D}8/\overline{\mathrm{D}8}$-brane.
Since baryon in this model could be equivalently represented by the
instanton configurations on the light-flavoured brane and the glueball
field is identified as the bulk gravitational waves, we solve the
classical equations of motion for the HL field with instanton solution
for the gauge fields. In the limitation of large $\lambda$ followed
by large $m_{H}$, we derive the mass formula of the soliton as the
onshell action of the flavour brane by taking account of the HL field
and bulk gravitational waves. Then following the collectivization
and quantization of the soliton in \cite{key-22,key-25,key-26}, the
effective Hamiltonian for the collective modes of heavy-flavoured
baryons is obtained which includes the interaction with glueball.
Afterwards, we examine the possible decay processes and compute the
associated decay rates with the effective Hamiltonian. We find these
decay rates are in agreement with the previous works by using this
model as in \cite{key-14,key-15,key-16,key-19} since they are proportional
to $\lambda^{-1}$. Then by comparing the quantum numbers of the baryonic
states with some experimental data and employing the identification
of baryonic states in \cite{key-22}, we find that one decay process
might be realistic and could be interpreted as the decay of baryonic
B-meson involving the glueball candidate $f_{0}\left(1710\right)$
as  discussed in \cite{key-8,key-9,key-10}. Noteworthily according
to lattice QCD  $f_{0}\left(1710\right)$ is an excited state in the
glueball candidates which is just consistent with that the glueball
state discussed in transition VIII is also an excitation.

As an improvement of \cite{key-19}, this work provides an alternative
way to investigate the interaction of glueball and heavy-flavoured
baryons in strongly coupling system through the holographic approach
of the underlying string theory. Although this approach is quite principal
and contains few parameters, it is actually valid in the large $N_{c}$
limit. So phenomenological theories or models are always needed as
a comparison with holography.

\section*{Acknowledgements}

I would like to thank Anton Rebhan, Josef Leutgeb and Chao Wu for
valuable comments and discussions. SWL is supported by the research
startup foundation of Dalian Maritime University in 2019.

\section*{Appendix}

\subsection*{A. The bulk supergravity and glueball dynamics in the WSS model}

The WSS model is based on the $\mathrm{AdS_{7}/CFT_{6}}$ correspondence
of $N_{c}$ M5-branes in string theory which can be reduced to $N_{c}$
D4-branes compactified on $S^{1}$ in 10d bulk. So taking the large
$N_{c}$ limit, the bulk dynamic is described by the 10d type IIA
supergravity action which is given as,

\begin{equation}
S_{\mathrm{IIA}}=\frac{1}{2k_{10}^{2}}\int d^{10}x\sqrt{-g}e^{-2\Phi}\left(\mathcal{R}+4\nabla_{M}\Phi\nabla^{M}\Phi-\frac{1}{2}\left|F_{4}\right|^{2}\right),\tag{A-1}\label{eq:A-1}
\end{equation}
where $\Phi$ denotes the dilaton field, $2k_{10}^{2}=16G_{10}/g_{s}^{2}=\left(2\pi\right)^{7}l_{s}^{8}$
. $\mathcal{R},G_{10}$ is 10d scalar curvature and Newton constant
respectively. $F_{4}=dC_{3}$ is the field strength of the Romand-Romand
(R-R) 3-form $C_{3}$. The geometrical solution for the bulk metric
is given as,

\begin{align}
ds^{2}= & \left(\frac{U}{R}\right)^{3/2}\left[\eta_{\mu\nu}dX^{\mu}dX^{\nu}+f\left(U\right)\left(dX^{4}\right)^{2}\right]+\left(\frac{R}{U}\right)^{3/2}\left[\frac{dU^{2}}{f\left(U\right)}+U^{2}d\Omega_{4}^{2}\right],\nonumber \\
f\left(U\right)= & 1-\frac{U_{KK}^{3}}{U^{3}},\ e^{\Phi}=\left(\frac{U}{R}\right)^{3/4},\ F_{4}=\frac{2\pi N_{c}}{V_{4}}\epsilon_{4},\ R^{3}=\pi g_{s}N_{c}l_{s}^{3},\tag{A-2}\label{eq:A-2}
\end{align}
with a periodic condition for $X^{4}$,

\begin{equation}
X^{4}\sim X^{4}+2\pi\delta X^{4},\ \delta X^{4}=\frac{1}{M_{KK}}.\tag{A-3}
\end{equation}
And the $r,z,Z$ coordinate used in the paper is defined as,

\begin{equation}
U^{3}=U_{KK}^{3}+U_{KK}z^{2},\ Z=\frac{z}{U_{KK}},\ 1+Z^{2}=\frac{r^{6}}{r_{KK}^{6}},\ U_{KK}=\frac{r_{KK}^{2}}{4R}.\tag{A-4}
\end{equation}
Note that $\epsilon_{4}$ represents a unit volume element on $S^{4}$.
$g_{s},l_{s}$ denotes the string coupling constant and the length
of string. The indices $\mu,\nu$ in (\ref{eq:A-2}) run from 0 to
3. Additionally we could define the QCD variables in terms of,

\begin{equation}
\lambda=g_{\mathrm{YM}}^{2}N_{c},\ g_{\mathrm{YM}}^{2}=2\pi g_{s}l_{s}M_{KK},\tag{A-5}
\end{equation}
where $g_{\mathrm{YM}},\lambda$ respectively denotes the Yang-Mills
and the 't Hooft coupling constant. 

In this model the glueball fields are identified as the gravitational
fluctuations to the bulk solution (\ref{eq:A-2}), thus we could rewrite
the metric as $G_{MN}\rightarrow G_{MN}^{\left(0\right)}+\delta G_{MN}$
in order to involve the glueball field. The 10d metric reduced from
11d supergravity with gravitational fluctuations is,

\begin{align}
g_{\mu\nu} & =\frac{r^{3}}{L^{3}}\left[\left(1+\frac{L^{2}}{2r^{2}}\delta G_{11,11}\right)\eta_{\mu\nu}+\frac{L^{2}}{r^{2}}\delta G_{\mu\nu}\right],\nonumber \\
g_{44} & =\frac{r^{3}f}{L^{3}}\left[1+\frac{L^{2}}{2r^{2}}\delta G_{11,11}+\frac{L^{2}}{r^{2}f}\delta G_{44}\right],\nonumber \\
g_{rr} & =\frac{L}{rf}\left(1+\frac{L^{2}}{2r^{2}}\delta G_{11,11}+\frac{r^{2}f}{L^{2}}\delta G_{rr}\right),\nonumber \\
g_{r\mu} & =\frac{r}{L}\delta G_{r\mu},\ \ g_{\Omega\Omega}=\frac{r}{L}\left(\frac{L}{2}\right)^{2}\left(1+\frac{L^{2}}{2r^{2}}\delta G_{11,11}\right),\tag{A-6}\label{eq:A-6}
\end{align}
with the dilaton,

\begin{equation}
e^{4\Phi/3}=\frac{r^{2}}{L^{2}}\left(1+\frac{L^{2}}{r^{2}}\delta G_{11,11}\right).\tag{A-7}\label{eq:A-7}
\end{equation}
Since different formulas of $\delta G_{MN}$ corresponds to various
glueball, in this paper we consider the following forms of $\delta G_{MN}$:

\subsubsection*{The exotic scalar glueball}

The exotic scalar glueball corresponds to the exotic polarizations
of the bulk graviton whose quantum number is $J^{CP}=0^{++}$. The
11d components of $\delta G_{MN}$ are given as ,

\begin{align}
\delta G_{44} & =-\frac{r^{2}}{L^{2}}f\left(r\right)H_{E}\left(r\right)G_{E}\left(x\right),\nonumber \\
\delta G_{\mu\nu} & =\frac{r^{2}}{L^{2}}H_{E}\left(r\right)\left[\frac{1}{4}\eta_{\mu\nu}-\left(\frac{1}{4}+\frac{3r_{KK}^{6}}{5r^{6}-2r_{KK}^{6}}\right)\frac{\partial_{\mu}\partial_{\nu}}{M_{E}^{2}}\right]G_{E}\left(x\right),\nonumber \\
\delta G_{11,11} & =\frac{r^{2}}{4L^{2}}H_{E}\left(r\right)G_{E}\left(x\right),\nonumber \\
\delta G_{rr} & =-\frac{L^{2}}{r^{2}}\frac{1}{f\left(r\right)}\frac{3r_{KK}^{6}}{5r^{6}-2r_{KK}^{6}}H_{E}\left(r\right)G_{E}\left(x\right),\nonumber \\
\delta G_{r\mu} & =\frac{90r^{7}r_{KK}^{6}}{M_{E}^{2}L^{2}\left(5r^{6}-2r_{KK}^{6}\right)^{2}}H_{E}\left(r\right)\partial_{\mu}G_{E}\left(x\right),\tag{A-8}\label{eq:A-8}
\end{align}
with the eigenvalue equation for function $H_{E}\left(r\right)$ as,

\begin{equation}
\frac{1}{r^{3}}\frac{d}{dr}\left[r\left(r^{6}-r_{KK}^{6}\right)\frac{d}{dr}H_{E}\left(r\right)\right]+\left[\frac{432r^{2}r_{KK}^{12}}{\left(5r^{6}-2r_{KK}^{6}\right)^{2}}+L^{4}M_{E}^{2}\right]H_{E}\left(r\right)=0.\tag{A-9}\label{eq:A-9}
\end{equation}
In 10d bulk the above components in (\ref{eq:A-8}) satisfy the asymptotics
$\delta G_{44}=-4\delta G_{11}=-4\delta G_{22}=-4\delta G_{33}=-4\delta G_{11,11}$
for $r\rightarrow\infty$. Plugging the solution (\ref{eq:A-2}) and
the fluctuations (\ref{eq:A-8}) with the eigenvalue equation (\ref{eq:A-9})
into the action (\ref{eq:A-1}), it leads to the kinetic term of the
exotic scalar glueball,

\begin{equation}
S_{G_{E}\left(x\right)}=-\frac{1}{2}\int d^{4}x\left[\left(\partial_{\mu}G_{E}\right)^{2}+M_{E}^{2}G_{E}^{2}\right],\tag{A-10}\label{eq:A-10}
\end{equation}
where the pre-factor in (\ref{eq:A-10}) has been normalized to $-1/2$
by choosing the boundary value of $H_{E}\left(r\right)$.

\subsubsection*{The dilatonic and tensor glueball}

The fluctuations of the metric,

\begin{align}
\delta G_{11,11} & =-3\frac{r^{2}}{L^{2}}H_{D}\left(r\right)G_{D}\left(x\right),\nonumber \\
\delta G_{\mu\nu} & =\frac{r^{2}}{L^{2}}H_{D}\left(r\right)\left[\eta^{\mu\nu}-\frac{\partial^{\mu}\partial^{\nu}}{M_{D}^{2}}\right]G_{D}\left(x\right),\tag{A-12}\label{eq:A-12}
\end{align}
corresponds to another mode of the scalar glueball $0^{++}$. We employ
``dilatonic'' for the upon mode since $\delta G_{11,11}$ reduces
to the 10d dilaton.

Besides the tensor glueball corresponds to the metric fluctuations
with a transverse traceless polarization whose quantum number is $J^{CP}=2^{++}$.
We can choose the following components of the graviton polarizations
as tensor glueball field,

\begin{equation}
\delta G_{\mu\nu}=-\frac{r^{2}}{L^{2}}H_{T}\left(r\right)T_{\mu\nu}\left(x\right),\tag{A-13}\label{eq:A-13}
\end{equation}
where $T_{\mu\nu}\equiv\mathcal{T}_{\mu\nu}G_{T}\left(x\right)$.
$\mathcal{T}_{\mu\nu}$ is a constant symmetric tensor satisfying
the normalization and traceless condition $\mathcal{T}_{\mu\nu}\mathcal{T}^{\mu\nu}=1,\eta^{\mu\nu}\mathcal{T}_{\mu\nu}=0$.
The functions $H_{D,T}\left(r\right)$ satisfies the eigenvalue equation,

\begin{equation}
\frac{1}{r^{3}}\frac{d}{dr}\left[r\left(r^{6}-r_{KK}^{6}\right)\frac{d}{dr}H_{D,T}\left(r\right)\right]+L^{4}M_{D,T}^{2}H_{D,T}\left(r\right)=0.\tag{A-14}\label{eq:A-14}
\end{equation}
We can also obtain the kinetic action of the dilatonic scalar and
tensor glueball as,

\begin{align}
S_{G_{D}\left(x\right)} & =-\frac{1}{2}\int d^{4}x\left[\left(\partial_{\mu}G_{D}\right)^{2}+M_{D}^{2}G_{D}^{2}\right],\nonumber \\
S_{T\left(x\right)} & =-\frac{1}{4}\int d^{4}x\left[T_{\mu\nu}\left(\partial^{2}-M_{T}^{2}\right)T^{\mu\nu}\right],\tag{A-15}\label{eq:A-15}
\end{align}
once the solution (\ref{eq:A-2}) and fluctuations (\ref{eq:A-12})
(\ref{eq:A-13}) with eigenvalue equation (\ref{eq:A-14}) are imposed
to the action (\ref{eq:A-1}) and the boundary value of $H_{D,T}$
has to been determined by the normalization conditions in (\ref{eq:A-15}).

\subsection*{B. The full Dp-brane action and the embedding of the probe branes}

\subsubsection*{The complete DBI action}

We give the complete formula of the Dp-brane here and it could also
be reviewed in many textbooks of string theory, Let us consider $D$
dimensional spacetime parametrized by $\left\{ X^{\mu}\right\} ,\mu=0,1...D-1$
with a stack of D$p$-branes. In this subsection, the indices $a,b=0,1...p$
and $i,j,k=p+1...D-1$ denote respectively the directions parallel
and vertical to the D$p$-branes. The complete bosonic action of a
D$p$-branes is,

\begin{equation}
S_{\mathrm{D}_{p}-\mathrm{branes}}=S_{\mathrm{DBI}}+S_{\mathrm{CS}},\tag{B-1}
\end{equation}
where \cite{key-27}

\begin{align}
S_{\mathrm{DBI}}= & -T_{p}\mathrm{STr}\int d^{p+1}\xi e^{-\Phi}\sqrt{-\det\left\{ \left[E_{ab}+E_{ai}\left(Q^{-1}-\delta\right)^{ij}E_{jb}+2\pi\alpha^{\prime}F_{ab}\right]Q_{\ j}^{i}\right\} },\nonumber \\
S_{\mathrm{CS}}= & \mu_{p}\sum_{n=0,1}\int_{\mathrm{D}_{p}\mathrm{-branes}}C_{p-2n+1}\wedge\frac{\left(B+2\pi\alpha^{\prime}F\right)^{n}}{n!},\nonumber \\
Q_{\ j}^{i}= & \delta^{i}\ j+2\pi\alpha^{\prime}\left[\varphi^{i},\varphi^{k}\right]E_{kj},\ E_{\mu\nu}=g_{\mu\nu}+B_{\mu\nu}.\tag{B-2}\label{eq:B-2}
\end{align}
We have denoted the metric of the $D$ dimensional spacetime and the
2-form field as $g_{\mu\nu},B_{\mu\nu}$ respectively. $F$ is the
gauge field strength defined on the D-brane and ``STr'' refers to
the ``symmetric trace''. We use $\varphi^{i}$ 's to represent the
transverse modes of the D$p$-branes which are given by the T-duality
relation $2\pi\alpha^{\prime}\varphi^{i}=X^{i}$. So the DBI action
in (\ref{eq:B-2}) could be expanded as, 

\begin{equation}
S_{\mathrm{BDI}}=-T_{p}\mathrm{Tr}\int d^{p+1}\xi e^{-\Phi}\sqrt{-g}\left[1+\frac{1}{4}\left(2\pi\alpha^{\prime}\right)^{2}F_{ab}F^{ab}+\frac{1}{2}D_{a}\varphi^{i}D_{a}\varphi^{i}+\frac{1}{4}\left[\varphi^{i},\varphi^{j}\right]^{2}\right]+\mathrm{high\ orders}.\tag{B-3}\label{eq:B-3}
\end{equation}
The 2-form field $B$ has been gauged away. The gauge field $A_{a}$
and scalar field $\varphi^{i}$ 's are all in the adjoint representation
of $U\left(N\right)$. Note that there is only one transverse coordinate
for the $\mathrm{D}8/\overline{\mathrm{D}8}$-brane which has been
defined as $\Psi\equiv\varphi^{9}$ in the main text.

\subsubsection*{Comments about the the probe branes and strings}

Here let us briefly outline the embedding of the probe $\mathrm{D}8/\overline{\mathrm{D}8}$-brane
and the HL string. Using the bulk metric (\ref{eq:A-2}), the induced
metric on the probe $\mathrm{D}8/\overline{\mathrm{D}8}$-branes is
obtained as,

\begin{equation}
ds_{\mathrm{D}8/\overline{\mathrm{D}8}}^{2}=\left(\frac{U}{R}\right)^{3/2}\left[f\left(U\right)+\left(\frac{R}{U}\right)^{3}\frac{U^{\prime2}}{f\left(U\right)}\right]\left(dX^{4}\right)^{2}+\left(\frac{U}{R}\right)^{3/2}\eta_{\mu\nu}dX^{\mu}dX^{\nu}+\left(\frac{R}{U}\right)^{3/2}U^{2}d\Omega_{4}^{2},\tag{B-4}\label{eq:B-4}
\end{equation}
where $U^{\prime}=\frac{dU}{dX^{4}}$. Then insert the metric (\ref{eq:B-4})
into the DBI action of $\mathrm{D}8/\overline{\mathrm{D}8}$-branes,
it yields the formula,

\begin{equation}
S_{\mathrm{D}8/\overline{\mathrm{D}8}}\propto\int d^{4}xdUU^{4}\left[f\left(U\right)+\left(\frac{R}{U}\right)^{3}\frac{U^{\prime2}}{f\left(U\right)}\right]^{1/2}.\tag{B-5}
\end{equation}
Hence we can obtain the equation of motion for the function $U\left(X^{4}\right)$
as,

\begin{equation}
\frac{d}{dX^{4}}\left(\frac{U^{4}f\left(U\right)}{\left[f\left(U\right)+\left(\frac{R}{U}\right)^{3}\frac{1}{f\left(U\right)}U^{\prime2}\right]^{1/2}}\right)=0.\tag{B-6}\label{eq:B-6}
\end{equation}
Using the boundary condition in \cite{key-13}, as $U\left(X^{4}=0\right)=U_{0}$
and $U^{\prime}\left(X^{4}=0\right)=0$, the generic solution for
(\ref{eq:B-6}) is computed as,

\begin{equation}
X^{4}\left(U\right)=E\left(U_{0}\right)\int_{U_{0}}^{U}dU\frac{\left(U\right)\left(\frac{R}{U}\right)^{3/2}}{f\left(U\right)\left[U^{8}f\left(U\right)-E^{2}\left(U_{0}\right)\right]^{1/2}},\tag{B-7}\label{eq:B-7}
\end{equation}
where $E\left(U_{0}\right)=U_{0}^{4}f^{1/2}\left(U_{0}\right)$ and
we have used $U_{0}$ to denotes the connected position of the $\mathrm{D}8/\overline{\mathrm{D}8}$-branes.
Afterwards let us further introduce the coordinates $\left(r,\Theta\right)$
and $\left(y,z\right)$ which satisfy,

\begin{align}
y=r\cos\Theta, & \ \ z=r\sin\Theta,\nonumber \\
U^{3}=U_{KK}^{3}+U_{KK}r^{2}, & \ \ \Theta=\frac{2\pi}{\beta}X^{4}=\frac{3}{2}\frac{U_{KK}^{1/2}}{R^{3/2}}.\tag{B-8}
\end{align}
In the standard WSS model, the probe $\mathrm{D}8/\overline{\mathrm{D}8}$-branes
are embedded at $\Theta=\pm\frac{1}{2}\pi$ respectively i.e. the
position of $y=0$, which exactly corresponds to the antipodal $\mathrm{D}8/\overline{\mathrm{D}8}$-branes
(blue) in Figure \ref{fig:1}. In this case, the solution for the
embedding function is $X^{4}\left(U\right)=\frac{1}{4}\beta$ and
$U_{0}=U_{KK}$. In addition, the (\ref{eq:B-7}) also allows the
non-antipodal solution if we choose $\Theta=\pm\Theta_{H}\neq\pm\frac{1}{2}\pi,U_{0}=U_{H}\neq U_{KK}$
which corresponds to the non-antipodal $\mathrm{D}8/\overline{\mathrm{D}8}$-branes
(red) in Figure \ref{fig:1}. On the other hand, while each endpoints
of the HL string could move along the flavoured branes, in our setup
it is stretched between the heavy- (non-antipodal) and light-flavoured
(antipodal) $\mathrm{D}8/\overline{\mathrm{D}8}$-branes. So it connects
the positions respectively on the heavy- and light-flavoured $\mathrm{D}8/\overline{\mathrm{D}8}$-branes
which are most close to each other and in the $U-X^{4}$ plane, they
are the positions of $\left(U_{KK},0\right)$ on the light-flavoured
branes and $\left(U_{H},0\right)$ on the heavy-flavoured branes.
And this is the configuration of the HL string with minimal length
i.e. the VEV.

\subsection*{C. The collective modes of baryon and its quantization}

As the $\mathrm{D4}^{\prime}$-brane is identified as baryon in the
WSS model, it is equivalent to the instanton configuration on the
D8-branes according to the string theory. So the dynamic of the $\mathrm{D}8/\overline{\mathrm{D}8}$-brane
is given by the Dirac-Born-Infield (DBI) action plus the Chern-Simons
(CS) action (\ref{eq:B-2}) while the baryonic $\mathrm{D4}^{\prime}$-brane
is identified as the instanton configuration of the gauge field strength
on the $\mathrm{D}8/\overline{\mathrm{D}8}$-brane. Altogether the
action of the flavours with baryons can be simplified as a 5d Yang-Mills
(YM) plus CS action by integrating over the $S^{4}$ which is given
as,

\begin{align}
S & =S_{\mathrm{YM}}+S_{\mathrm{CS}}.\nonumber \\
S_{\mathrm{YM}} & =-\kappa\mathrm{Tr}\int d^{4}xdze^{-\Phi}\sqrt{-g}g^{ab}g^{cd}\mathcal{F}_{ac}\mathcal{F}_{bd},\nonumber \\
S_{\mathrm{CS}} & =\frac{N_{c}}{24\pi}\mathrm{Tr}\int d^{4}xdz\left(\mathcal{A}\mathcal{F}^{2}-\frac{1}{2}\mathcal{A}^{3}\mathcal{F}-\frac{1}{10}\mathcal{A}^{5}\right),\tag{C-1}\label{eq:C-1}
\end{align}
where the indices $\alpha,\beta$ run over $X^{\mu}$ and $z$. Particularly
in the situation of two flavours i.e. $N_{f}=2$, the classical instanton
configuration could be adopted as the Belavin-Polyakov-Schwarz-Tyupkin
(BPST) solution which is given as,

\begin{align}
\mathcal{A}_{M}= & -\bar{\sigma}_{MN}\frac{x^{N}}{x^{2}+\rho^{2}},\ M,N=1,2,3,z,\nonumber \\
\mathcal{A}_{0}= & -\frac{i}{8\pi^{2}ab^{3/2}x^{2}}\left[1-\frac{\rho^{4}}{\left(x^{2}+\rho^{2}\right)^{2}}\right],\tag{C-2}\label{eq:C-2}
\end{align}
where $\mathcal{A}$ is $U\left(2\right)$ and $\mathcal{A}_{0}$
is $U\left(1\right)$ gauge field . The gauge field strength is defined
as $\mathcal{F}=d\mathcal{A}+\left[\mathcal{A},\mathcal{A}\right]$\footnote{In our notation, $\mathcal{A}$ is anti-Hermitian which means $\mathcal{A}^{\dagger}=-\mathcal{A}$. }.
And $x^{2}=\left(x^{M}-X^{M}\right)^{2}$, $X^{M}$ 's are constants.
Since the instanton size $\rho$ is of order $\lambda^{-1/2}$, it
would be convenient to employ the rescaling,

\begin{equation}
\left(x^{0},x^{M}\right)\rightarrow\left(x^{0},\lambda^{-1/2}x^{M}\right),\ \left(\mathcal{A}_{0},\mathcal{A}_{M}\right)\rightarrow\left(\mathcal{A}_{0},\lambda^{1/2}\mathcal{A}_{M}\right),\tag{C-3}\label{eq:C-3}
\end{equation}
in order to obtain the explicit dependence of $\lambda$ in the actions
in (\ref{eq:C-1}). Inserting (\ref{eq:C-2}) into the rescaled gauge
field $\mathcal{A}$, the mass $M$ of the classical soliton could
be evaluated by $S_{cl}^{\mathrm{onshell}}=-\int dtM$. Afterwards
the baryon states could be identified as Skyrmions so that the characteristics
of baryon are reflected by their collective modes. Therefore we could
quantize the classical soliton in the moduli space to obtain the baryon
spectrum.

In the large $\lambda$ limit, the topology of the moduli space for
$N_{f}=2$ case is given as $\mathbb{R}^{4}\times\mathbb{R}^{4}/\mathbb{Z}_{2}$
since the contribution of $\mathcal{O}\left(\lambda^{-1}\right)$
could be neglected. The the collective coordinates $\left\{ X^{M}\right\} $
parameterize the first $\mathbb{R}^{4}$ while the size $\rho$ and
the $SU(2)$ orientation of the instanton parameterize $\mathbb{R}^{4}/\mathbb{Z}_{2}$
. Let us denote the $SU(2)$ orientation as $a_{I}=\frac{y_{I}}{\rho},\ I=1,2,3,4$
with the normalization $\sum_{I=1}^{4}a_{I}^{2}=1$ so that the size
of the instanton satisfies $\rho=\sqrt{y_{1}^{2}+...y_{4}^{2}}$.
The quantization procedures of the Lagrangian for the collective coordinates
follows those in Ref. Specifically we need to assume that the moduli
of the solution is time-dependent. Thus the gauge transformation also
becomes time-dependent as,

\begin{align}
\mathcal{A}_{M} & \rightarrow V\left(\mathcal{A}_{M}^{cl}-i\partial_{M}\right)V^{-1},\nonumber \\
\mathcal{F}_{MN} & \rightarrow V\mathcal{F}_{MN}^{cl}V^{-1},\ F_{0M}\rightarrow V\left(\dot{X}^{\alpha}\partial_{\alpha}\mathcal{A}_{M}^{cl}-D_{M}^{cl}\Phi\right)V^{-1},\tag{C-4}\label{eq:C-4}
\end{align}
The Lagrangian of the collective coordinates in such a moduli space
takes the form as,

\begin{equation}
L=\frac{m_{X}}{2}\mathcal{G}_{rs}\dot{\mathcal{X}^{s}}\dot{\mathcal{X}^{r}}-U\left(\mathcal{X}^{s}\right)+\mathcal{O}\left(\lambda^{-1}\right),\tag{C-5}\label{eq:C-5}
\end{equation}
where $\mathcal{X}^{s}=\left\{ X^{M},a_{I}\right\} $. The the kinetic
term in (\ref{eq:C-5}) corresponds to the line element of the moduli
space while the potential corresponds to the onshell action of the
soliton adopting the time-dependent gauge transformation,

\[
S_{\mathrm{D}8/\overline{\mathrm{D}8}}^{\mathrm{onshell}}\simeq S_{YM+CS}^{\mathrm{onshell}}=-\int dtU(\mathcal{X}^{s}).\tag{C-6}
\]
Using the solution (\ref{eq:C-2}), the above integral is easy to
calculate in the case of pure light flavours while it becomes quite
difficult if the heavy flavour is involved. Without loss of generality,
let us consider the large $\lambda$ limit followed by heavy mass
limit of the heavy flavour. Hence the dimensionless quantized Hamiltonian
corresponding to (\ref{eq:C-5}) for the collective modes is calculated
as,

\noindent 
\begin{align}
H & =M_{0}+H_{y}+H_{Z}+\mathcal{O}\left(\lambda^{-1}m_{H}^{0}\right),\nonumber \\
H_{y} & =-\frac{1}{2m_{y}}\sum_{I=1}^{4}\frac{\partial^{2}}{\partial y_{I}^{2}}+\frac{1}{2}m_{y}\omega_{y}^{2}\rho^{2}+\frac{\mathcal{Q}}{\rho^{2}},\nonumber \\
H_{Z} & =-\frac{1}{2m_{Z}}\frac{\partial^{2}}{\partial Z^{2}}+\frac{1}{2}m_{Z}\omega_{Z}^{2}Z^{2},\tag{C-7}\label{eq:C-7}
\end{align}
where,

\begin{align}
M_{0} & =8\pi^{2}\kappa,~\ \omega_{Z}^{2}=\frac{2}{3},~~\omega_{\rho}^{2}=\frac{1}{6},\ \ \kappa=\frac{\lambda N_{c}}{216\pi^{3}},\nonumber \\
\mathcal{Q} & =Q_{L}+Q_{H},\ \ Q_{L}=\frac{N_{c}}{40\pi^{2}a},\ \ Q_{H}=\frac{N_{Q}}{8\pi^{2}a}\left(\frac{N_{Q}}{3N_{c}}-\frac{3}{4}\right).\tag{C-8}
\end{align}
The value of $\mathcal{Q}$ corresponds to the situation of a baryonic
bound state consisting of $N_{Q}$ heavy flavoured quarks. The eigenfunctions
and mass spectrum of (\ref{eq:C-7}) can be evaluated by solving its
Schrodinger equation, respectively they are obtained as\footnote{$l$ and $\tilde{l}$ are related as $\tilde{l}=-1+\sqrt{\left(l+1\right)^{2}+2m_{y}\mathcal{Q}}$. },

\begin{align}
\psi(y_{I}) & =R(\rho)T^{(l)}(a_{I}),\ R(\rho)=e^{-\frac{m_{y}\omega_{\rho}}{2}\rho^{2}}\rho^{\tilde{l}}Hypergeometric_{1}F_{1}\left(-n_{\rho},\tilde{l}+2;m_{y}\omega_{\rho}\rho^{2}\right),\nonumber \\
E\left(l,n_{\rho},n_{z}\right) & =\omega_{\rho}\left(\tilde{l}+2n_{\rho}+2\right)=\sqrt{\frac{(l+1)^{2}}{6}+\frac{640}{3}a^{2}\pi^{4}Q^{2}}+\frac{2\left(n_{\rho}+n_{z}\right)+2}{\sqrt{6}}.\tag{C-9}\label{eq:C-9}
\end{align}
Notice that $T^{(l)}(a_{I})$ satisfies $\nabla_{S^{3}}^{2}T^{(l)}=-l(l+2)T^{(l)}$
which is the function of the spherical part because $H_{y}$ can be
written with the radial coordinate $\rho$ as,

\begin{equation}
H_{y}=-\frac{1}{2m_{y}}\left[\frac{1}{\rho^{3}}\partial_{\rho}(\rho^{3}\partial_{\rho})+\frac{1}{\rho^{2}}\left(\nabla_{S^{3}}^{2}-2m_{y}\mathcal{Q}\right)\right]+\frac{1}{2}m_{y}\omega_{\rho}^{2}\rho^{2}.\tag{C-10}\label{eq:C-10}
\end{equation}

\subsection*{D. Explicit formulas of $\mathcal{L}_{1/2,0,-1/2}^{E,D,T}$ and $\mathcal{L}_{\Psi}^{E,D,T}$}

Here we collect the explicit formulas of $\mathcal{L}_{1/2,0,-1/2}^{E,D,T}$
and $\mathcal{L}_{\Psi}^{E,D,T}$. For the exotic scalar glueball,

\begin{align}
\mathcal{L}_{1/2}^{E}= & \frac{1}{M_{KK}}\bigg[-\frac{5}{4M_{E}^{2}}\partial^{i}\partial^{j}G_{E}\boldsymbol{\mathcal{F}}_{ik}\boldsymbol{\mathcal{F}}_{j}^{\ k}+\frac{3}{16}G_{E}\boldsymbol{\mathcal{F}}_{ij}\boldsymbol{\mathcal{F}}^{ij}+\frac{5}{16M_{E}^{2}}\partial^{2}G_{E}\boldsymbol{\mathcal{F}}_{ij}\boldsymbol{\mathcal{F}}^{ij}\nonumber \\
 & -\frac{5}{4M_{E}^{2}}\partial^{i}\partial^{j}G_{E}\boldsymbol{\mathcal{F}}_{iZ}\boldsymbol{\mathcal{F}}_{jZ}-\frac{7}{8}\eta^{ij}G_{E}\boldsymbol{\mathcal{F}}_{iZ}\boldsymbol{\mathcal{F}}_{jZ}+\frac{5}{8M_{E}^{2}}\partial^{2}G_{E}\eta^{ij}\boldsymbol{\mathcal{F}}_{iZ}\boldsymbol{\mathcal{F}}_{jZ}\bigg],\nonumber \\
\mathcal{L}_{0}^{E}= & \frac{1}{M_{E}^{2}}\bigg[\frac{20}{3}\partial^{k}G_{E}\eta^{ij}Z\boldsymbol{\mathcal{F}}_{ik}\boldsymbol{\mathcal{F}}_{jZ}-\frac{5}{2M_{KK}}\partial^{0}\partial^{k}G_{E}\eta^{ij}\boldsymbol{\mathcal{F}}_{ik}\boldsymbol{\mathcal{F}}_{j0}-\frac{5}{2M_{KK}}\partial^{0}\partial^{i}G_{E}\boldsymbol{\mathcal{F}}_{Zi}\boldsymbol{\mathcal{F}}_{Z0}\bigg],\nonumber \\
\mathcal{L}_{-1/2}^{E}= & \frac{Z^{2}}{M_{KK}}\bigg[\frac{5}{16M_{KK}^{2}}\partial^{i}\partial^{j}G_{E}\boldsymbol{\mathcal{F}}_{ik}\boldsymbol{\mathcal{F}}_{j}^{\ k}+\frac{15}{4M_{E}^{2}}\partial^{i}\partial^{j}G_{E}\boldsymbol{\mathcal{F}}_{ik}\boldsymbol{\mathcal{F}}_{j}^{\ k}-\frac{5}{64M_{KK}^{2}}\partial^{2}G_{E}\boldsymbol{\mathcal{F}}_{ij}\boldsymbol{\mathcal{F}}^{ij}\nonumber \\
 & -\frac{3M_{E}^{2}}{64M_{KK}^{2}}G_{E}\boldsymbol{\mathcal{F}}_{ij}\boldsymbol{\mathcal{F}}^{ij}-\frac{35}{48}G_{E}\boldsymbol{\mathcal{F}}_{ij}\boldsymbol{\mathcal{F}}^{ij}-\frac{15}{16M_{E}^{2}}\partial^{2}G_{E}\boldsymbol{\mathcal{F}}_{ij}\boldsymbol{\mathcal{F}}^{ij}+\frac{5}{16}\partial^{i}\partial^{j}G_{E}\boldsymbol{\mathcal{F}}_{iZ}\boldsymbol{\mathcal{F}}_{jZ}\nonumber \\
 & +\frac{25}{12}\frac{M_{KK}^{2}}{M_{E}^{2}}\partial^{i}\partial^{j}G_{E}\boldsymbol{\mathcal{F}}_{iZ}\boldsymbol{\mathcal{F}}_{jZ}-\frac{5}{32}\partial^{2}G_{E}\eta^{ij}\boldsymbol{\mathcal{F}}_{iZ}\boldsymbol{\mathcal{F}}_{jZ}+\frac{7}{32}M_{E}^{2}G_{E}\eta^{ij}\boldsymbol{\mathcal{F}}_{iZ}\boldsymbol{\mathcal{F}}_{jZ}\nonumber \\
 & +\frac{9}{8}G_{E}M_{KK}^{2}\eta^{ij}\boldsymbol{\mathcal{F}}_{iZ}\boldsymbol{\mathcal{F}}_{jZ}-\frac{25}{24}\frac{M_{KK}^{2}}{M_{E}^{2}}\partial^{2}G_{E}\eta^{ij}\boldsymbol{\mathcal{F}}_{iZ}\boldsymbol{\mathcal{F}}_{jZ}\bigg]+\frac{1}{M_{E}^{2}M_{KK}}\bigg[\frac{5}{4}\partial^{i}\partial^{j}G_{E}\boldsymbol{\mathcal{F}}_{i0}\boldsymbol{\mathcal{F}}_{j0}\nonumber \\
 & -\frac{3}{8}M_{E}^{2}G_{E}\eta^{ij}\boldsymbol{\mathcal{F}}_{i0}\boldsymbol{\mathcal{F}}_{j0}-\frac{5}{8}\partial^{2}G_{E}\eta^{ij}\boldsymbol{\mathcal{F}}_{i0}\boldsymbol{\mathcal{F}}_{j0}-\frac{5}{4}\partial^{0}\partial^{0}G_{E}\eta^{ij}\boldsymbol{\mathcal{F}}_{i0}\boldsymbol{\mathcal{F}}_{j0}\bigg]\nonumber \\
 & +\frac{1}{M_{E}^{2}M_{KK}}\left[\frac{7}{8}G_{E}M_{E}^{2}\boldsymbol{\mathcal{F}}_{Z0}\boldsymbol{\mathcal{F}}_{Z0}-\frac{5}{8}\partial^{2}G_{E}\boldsymbol{\mathcal{F}}_{Z0}\boldsymbol{\mathcal{F}}_{Z0}-\frac{5}{4}\partial^{0}\partial^{0}G_{E}\boldsymbol{\mathcal{F}}_{Z0}\boldsymbol{\mathcal{F}}_{Z0}\right]\nonumber \\
 & -\frac{20}{3M_{E}^{2}}Z\partial^{0}G_{E}\eta^{ij}\boldsymbol{\mathcal{F}}_{jZ}\boldsymbol{\mathcal{F}}_{i0},\nonumber \\
\mathcal{L}_{\Psi}^{E}= & -v^{2}\frac{\left(N_{f}+1\right)^{2}}{N_{f}^{2}}\bigg[-\frac{5}{12M_{E}^{2}M_{KK}}\partial^{i}\partial^{j}G_{E}\Phi_{i}^{\dagger}\Phi_{j}+\frac{5}{24M_{E}^{2}M_{KK}}\partial^{2}G_{E}\delta^{ij}\Phi_{i}^{\dagger}\Phi_{j}\nonumber \\
 & -\frac{5}{12M_{KK}}G_{E}\Phi_{Z}^{\dagger}\Phi_{Z}+\frac{5}{24M_{E}^{2}M_{KK}}\partial^{2}G_{E}\Phi_{Z}^{\dagger}\Phi_{Z}\bigg].\tag{D-1}\label{eq:D-1}
\end{align}
For the dilatonic scalar glueball,

\begin{align}
\mathcal{L}_{1/2}^{D}= & -\frac{\partial^{i}\partial^{j}G_{D}}{M_{D}^{2}M_{KK}}\boldsymbol{\mathcal{F}}_{ik}\boldsymbol{\mathcal{F}}_{j}^{\ k}+\frac{3G_{D}}{4M_{KK}}\boldsymbol{\mathcal{F}}_{ij}\boldsymbol{\mathcal{F}}^{ij}+\frac{\partial^{2}G_{D}}{4M_{D}^{2}M_{KK}}\boldsymbol{\mathcal{F}}_{ij}\boldsymbol{\mathcal{F}}^{ij}\nonumber \\
 & -\frac{\partial^{i}\partial^{j}G_{D}}{M_{D}^{2}M_{KK}}\boldsymbol{\mathcal{F}}_{iZ}\boldsymbol{\mathcal{F}}_{jZ}+\frac{1}{2}G_{D}M_{KK}^{-1}\boldsymbol{\mathcal{F}}_{Zi}\boldsymbol{\mathcal{F}}_{Z}^{\ i}+\frac{\partial^{2}G_{D}}{2M_{D}^{2}M_{KK}}\boldsymbol{\mathcal{F}}_{Zi}\boldsymbol{\mathcal{F}}_{Z}^{\ i},\nonumber \\
\mathcal{L}_{-1/2}^{D}= & \frac{\partial^{i}\partial^{j}G_{D}}{4M_{KK}^{3}}Z^{2}\boldsymbol{\mathcal{F}}_{ik}\boldsymbol{\mathcal{F}}_{j}^{\ k}+\frac{\partial^{i}\partial^{j}G_{D}}{3M_{D}^{2}M_{KK}}Z^{2}\boldsymbol{\mathcal{F}}_{ik}\boldsymbol{\mathcal{F}}_{j}^{\ k}-\frac{\partial^{2}G_{D}}{16M_{KK}^{3}}Z^{2}\boldsymbol{\mathcal{F}}_{ij}\boldsymbol{\mathcal{F}}^{ij}\nonumber \\
 & -\frac{3G_{D}M_{D}^{2}}{16M_{KK}^{3}}Z^{2}\boldsymbol{\mathcal{F}}_{ij}\boldsymbol{\mathcal{F}}^{ij}-\frac{G_{D}}{4M_{KK}}Z^{2}\boldsymbol{\mathcal{F}}_{ij}\boldsymbol{\mathcal{F}}^{ij}-\frac{\partial^{2}G_{D}}{12M_{D}^{2}M_{KK}}Z^{2}\boldsymbol{\mathcal{F}}_{ij}\boldsymbol{\mathcal{F}}^{ij}\nonumber \\
 & +\frac{\partial^{i}\partial^{j}G_{D}}{4M_{KK}^{3}}Z^{2}\boldsymbol{\mathcal{F}}_{iZ}\boldsymbol{\mathcal{F}}_{jZ}-\frac{\partial^{i}\partial^{j}G_{D}Z^{2}}{M_{D}^{2}M_{KK}}\boldsymbol{\mathcal{F}}_{iZ}\boldsymbol{\mathcal{F}}_{jZ}-\frac{\partial^{2}G_{D}Z^{2}}{8M_{KK}^{3}}\boldsymbol{\mathcal{F}}_{Zi}\boldsymbol{\mathcal{F}}_{Z}^{\ i}\nonumber \\
 & -\frac{G_{D}M_{D}^{2}Z^{2}}{8M_{KK}^{3}}\boldsymbol{\mathcal{F}}_{Zi}\boldsymbol{\mathcal{F}}_{Z}^{\ i}+\frac{1}{2}G_{D}M_{KK}^{-1}Z^{2}\boldsymbol{\mathcal{F}}_{Zi}\boldsymbol{\mathcal{F}}_{Z}^{\ i}+\frac{\partial^{2}G_{D}Z^{2}}{2M_{D}^{2}M_{KK}}\boldsymbol{\mathcal{F}}_{Zi}\boldsymbol{\mathcal{F}}_{Z}^{\ i}\nonumber \\
 & +\frac{\partial^{i}\partial^{j}G_{D}}{M_{D}^{2}M_{KK}}\boldsymbol{\mathcal{F}}_{i0}\boldsymbol{\mathcal{F}}_{j0}-\frac{3}{2}\frac{G_{D}}{M_{KK}}\boldsymbol{\mathcal{F}}_{0i}\boldsymbol{\mathcal{F}}_{0}^{\ i}-\frac{\partial^{2}G_{D}}{2M_{D}^{2}M_{KK}}\boldsymbol{\mathcal{F}}_{0i}\boldsymbol{\mathcal{F}}_{0}^{\ i}-\frac{\partial^{0}\partial^{0}G_{D}}{M_{D}^{2}M_{KK}}\boldsymbol{\mathcal{F}}_{0i}\boldsymbol{\mathcal{F}}_{0}^{\ i}\nonumber \\
 & -\frac{1}{2}G_{D}M_{KK}^{-1}\boldsymbol{\mathcal{F}}_{0Z}^{2}-\frac{\partial^{2}G_{D}}{2M_{D}^{2}M_{KK}}\boldsymbol{\mathcal{F}}_{0Z}^{2}-\frac{\partial^{0}\partial^{0}G_{D}}{M_{D}^{2}M_{KK}}\boldsymbol{\mathcal{F}}_{0Z}^{2},\mathcal{L}_{0}^{D}=0,\nonumber \\
\frac{\mathcal{L}_{\Psi}^{D}}{a\mathcal{C}_{D}}= & v^{2}\frac{\left(N_{f}+1\right)^{2}}{N_{f}^{2}}\bigg[-\frac{\partial^{i}\partial^{j}G_{D}}{3M_{D}^{2}M_{KK}}\Phi_{i}^{\dagger}\Phi_{j}+\frac{2G_{D}}{3M_{KK}}\eta^{ij}\Phi_{i}^{\dagger}\Phi_{j}\nonumber \\
 & +\frac{\partial^{2}G_{D}}{6M_{D}^{2}M_{KK}}\eta^{ij}\Phi_{i}^{\dagger}\Phi_{j}+\frac{G_{D}}{3M_{KK}}\Phi_{Z}^{\dagger}\Phi_{Z}+\frac{\partial^{2}G_{D}}{6M_{D}^{2}M_{KK}}\Phi_{Z}^{\dagger}\Phi_{Z}\bigg].\tag{D-2}\label{eq:D-2}
\end{align}
For the tensor glueball

\begin{align}
\mathcal{L}_{1/2}^{T}= & -\frac{T^{kl}}{M_{KK}}\eta^{ij}\boldsymbol{\mathcal{F}}_{ik}\boldsymbol{\mathcal{F}}_{jl}-\frac{T^{ij}}{M_{KK}}\boldsymbol{\mathcal{F}}_{iZ}\boldsymbol{\mathcal{F}}_{jZ},\nonumber \\
\mathcal{L}_{0}^{T}= & -\frac{2T^{0k}}{M_{KK}}\eta^{ij}\boldsymbol{\mathcal{F}}_{ik}\boldsymbol{\mathcal{F}}_{j0}-\frac{2T^{0i}}{M_{KK}}\boldsymbol{\mathcal{F}}_{Zi}\boldsymbol{\mathcal{F}}_{Z0},\nonumber \\
\mathcal{L}_{-1/2}^{T}= & \frac{T^{kl}\eta^{ij}}{3M_{KK}}Z^{2}\boldsymbol{\mathcal{F}}_{ik}\boldsymbol{\mathcal{F}}_{jl}+\frac{M_{T}^{2}T^{kl}\eta^{ij}}{4M_{KK}^{3}}Z^{2}\boldsymbol{\mathcal{F}}_{ik}\boldsymbol{\mathcal{F}}_{jl}-M_{KK}T^{ij}Z^{2}\boldsymbol{\mathcal{F}}_{iZ}\boldsymbol{\mathcal{F}}_{jZ}\nonumber \\
 & +\frac{M_{T}^{2}}{4M_{KK}}T^{ij}Z^{2}\boldsymbol{\mathcal{F}}_{iZ}\boldsymbol{\mathcal{F}}_{jZ}+\frac{T^{ij}}{M_{KK}}\boldsymbol{\mathcal{F}}_{i0}\boldsymbol{\mathcal{F}}_{j0}-\frac{T^{00}}{M_{KK}}\eta^{ij}\boldsymbol{\mathcal{F}}_{i0}\boldsymbol{\mathcal{F}}_{j0}\nonumber \\
 & -\frac{T^{00}}{M_{KK}}\boldsymbol{\mathcal{F}}_{Z0}\boldsymbol{\mathcal{F}}_{Z0},\nonumber \\
\mathcal{L}_{\Psi}^{T}= & -\frac{\left(N_{f}+1\right)^{2}}{3N_{f}^{2}M_{KK}}v^{2}T^{ij}\Phi_{i}^{\dagger}\Phi_{i}.\tag{D-3}\label{eq:D-3}
\end{align}
We assume the glueball field is onshell so that $G_{E,D,T}$ could
be chosen as $G_{E,D,T}=\frac{1}{2}\left(e^{-iM_{E,D,T}t}+\mathrm{c.c}\right)$
in the rest frame of the glueball, hence we have $\partial^{i}G_{E,D,T}=0,\partial_{\mu}\partial^{\mu}G_{E,D,T}=M_{E,D,T}^{2}G_{E,D,T}$
which could greatly simplify (\ref{eq:D-1}) (\ref{eq:D-2}) (\ref{eq:D-3}).
Since the $\mathcal{L}_{\Psi}^{E,D,T}$ refers to the mass term of
the HL field, the mass of the heavy quarks $m_{H}$ must be related
to the separation of the flavour branes i.e. the VEV of $\Psi$. In
the heavy quark limit, the explicit relation is given as \cite{key-24,key-25,key-26},

\begin{align}
m_{H} & =\frac{1}{\pi l_{s}^{2}}\lim_{z_{H}\rightarrow\infty}\int_{0}^{z_{H}}dz\sqrt{-g_{00}g_{zz}},\nonumber \\
 & \simeq\frac{1}{\pi l_{s}^{2}}U_{KK}^{1/3}z_{H}^{2/3}+\mathcal{O}\left(z_{H}^{0}\right).\nonumber \\
 & \equiv\frac{1}{\sqrt{6}}\frac{N_{f}+1}{N_{f}}v,\tag{D-4}
\end{align}
where $z_{H}$ refers to the position $U=U_{H}$. Then we further
collect the terms of $\mathcal{O}\left(m_{H}^{2}\right)$ and $\mathcal{O}\left(m_{H}\right)$
then integral out the part of $z$, it finally leads to the formulas
in (\ref{eq:14}).


\begin{thebibliography}{10}
\bibitem{key-1}H. Fritzsch and M. Gell-Mann, ``Current algebra:
Quarks and what else?'', eConf C720906V2 (1972) 135\textendash 165,
{[}hep-ph/0208010{]}. 

\bibitem[2]{key-2}H. Fritzsch and P. Minkowski, ``\textgreek{Y}
Resonances, Gluons and the Zweig Rule'', Nuovo Cim. A30 (1975) 393. 

\bibitem[3]{key-3}R. Jaffe and K. Johnson, ``Unconventional States
of Confined Quarks and Gluons'', Phys.Lett. B60 (1976) 201.

\bibitem[4]{key-4}C. J. Morningstar and M. J. Peardon, The Glueball
spectrum from an anisotropic lattice study, Phys.Rev. D60 (1999) 034509,
{[}hep-lat/9901004{]}.

\bibitem[5]{key-5}Y. Chen, A. Alexandru, S. Dong, T. Draper, I. Horvath,
et al., Glueball spectrum and matrix elements on anisotropic lattices,
Phys.Rev. D73 (2006) 014516, {[}hep-lat/0510074{]}.

\bibitem[6]{key-6}H. B. Meyer and M. J. Teper, ``Glueball Regge
trajectories and the pomeron: A Lattice study'', Phys. Lett. B605
(2005) 344\textendash 354, {[}hep-ph/0409183{]}.

\bibitem[7]{key-7}E. Gregory, A. Irving, B. Lucini, C. McNeile, A.
Rago, C. Richards, E. Rinaldi, ``Towards the glueball spectrum from
unquenched lattice QCD'', JHEP10 (2012) 170, {[}arXiv:1208.1858{]}.

\bibitem[8]{key-8}Y.K. Hsiao, C.Q. Geng, \textquotedblleft Identifying
Glueball at 3.02 GeV in Baryonic B Decays\textquotedblright , Phys.
Lett. B 727 (2013) 168-171, {[}arXiv:1302.3331{]}.

\bibitem[9]{key-9}Xiao-Gang He, Tzu-Chiang Yuan, ``Glueball Production
via Gluonic Penguin B Decays'', Eur.Phys.J. C75 (2015) no.3, 136,
{[}arXiv:1503.03577{]}.

\bibitem[10]{key-10}Xian-Wei Kang, Tao Luo, Yi Zhang, Ling-Yun Dai,
Chao Wang, ``Semileptonic B and Bs decays involving scalar and axial-vector
mesons'', Eur.Phys.J. C78 (2018) no.11, 909, {[}arXiv:1808.02432{]}.

\bibitem[11]{key-11}K. Hashimoto, C.-I. Tan, and S. Terashima, Glueball
decay in holographic QCD, Phys.Rev. D77 (2008) 086001, {[}arXiv:0709.2208{]}.

\bibitem[12]{key-12}E. Witten, Anti-de Sitter space, thermal phase
transition, and confinement in gauge theories, Adv.Theor.Math.Phys.
2 (1998) 505\textendash 532, {[}hep-th/9803131{]}.

\bibitem[13]{key-13}T. Sakai and S. Sugimoto, Low energy hadron physics
in holographic QCD, Prog.Theor.Phys. 113 (2005) 843\textendash 882,
{[}hep-th/0412141{]}.

\bibitem[14]{key-14}Frederic Brünner, Denis Parganlija, Anton Rebhan,
``Glueball Decay Rates in the Witten-Sakai-Sugimoto Model'', Phys.Rev.
D91 (2015) no.10, 106002, (Erratum:) Phys.Rev. D93 (2016) no.10, 109903,
{[}arXiv:1501.07906{]}.

\bibitem[15]{key-15}Si-wen Li, ``The interaction of glueball and
heavy-light flavoured meson in holographic QCD'', {[}arXiv:1809.10379{]}.

\bibitem[16]{key-16}Frederic Brünner, Josef Leutgeb, Anton Rebhan,
``A broad pseudovector glueball from holographic QCD'', {[}arXiv:1807.10164{]}.

\bibitem[17]{key-17}N. R. Constable and R. C. Myers, \textquotedblleft Spin-two
glueballs, positive energy theorems and the AdS/CFT correspondence\textquotedblright ,
JHEP 9910 (1999) 037, {[}arXiv:hep-th/9908175{]}.

\bibitem[18]{key-18}R. C. Brower, S. D. Mathur, and C.-I. Tan, \textquotedblleft Glueball
spectrum for QCD from AdS supergravity duality\textquotedblright ,
Nucl.Phys. B587 (2000) 249\textendash 276, {[}arXiv:hep-th/0003115{]}.

\bibitem[19]{key-19}Si-wen Li, ``Glueball-baryon interactions in
holographic QCD'', Phys.Lett. B773 (2017) 142-149, {[}arXiv:1509.06914{]}.

\bibitem[20]{key-20}Edward Witten, ``Baryons and branes in anti-de
Sitter space'', JHEP 9807 (1998) 006, {[}hep-th/9805112{]}.

\bibitem[21]{key-21}Tong, David, ``TASI lectures on solitons: Instantons,
monopoles, vortices and kinks'' (2005), {[}hep-th/0509216{]}.

\bibitem[22]{key-22}Hiroyuki Hata, Tadakatsu Sakai, Shigeki Sugimoto,
Shinichiro Yamato, ``Baryons from instantons in holographic QCD'',
Prog.Theor.Phys.117:1157,2007, {[}arXiv:hep-th/0701280{]}.

\bibitem[23]{key-23}Y. Liu, I. Zahed, Holographic heavy-light chiral
effective action. Phys. Rev. D 95, 056022, {[}arXiv:1611.03757{]}. 

\bibitem[24]{key-24}Y. Liu, I. Zahed, Heavy-light mesons in chiral
AdS/QCD. Phys. Lett. B. {[}arXiv:1611.04400{]}.

\bibitem[25]{key-25}Yizhuang Liu, Ismail Zahed, ``Heavy Baryons
and their Exotics from Instantons in Holographic QCD'', Phys.Rev.
D95 (2017) no.11, 116012, {[}arXiv:1704.03412{]}.

\bibitem[26]{key-26}Si-wen Li, ``Holographic heavy-baryons in the
Witten-Sakai-Sugimoto model with the D0-D4 background'', Phys. Rev.
D 96, 106018 (2017), {[}arXiv:1707.06439{]}.

\bibitem[27]{key-27}Katrin Becker, Melanie Becker, John H. Schwarz,
``String theory and M-theory, A Modern Introduction'', Cambridge
University Press, 2007.

\bibitem[28]{key-28}R.C. Myers, ``Dielectric-Branes'', JHEP 9912,
022 (1999), {[}arXiv:hep-th/9910053{]}.
\end{thebibliography}
\end{document}